\documentclass{article}
\usepackage{chicago}
\usepackage{amssymb}
\usepackage{tikz}
\usetikzlibrary{decorations.pathmorphing}

\newcommand{\AdS}{\mbox{AdS}}

\newcommand{\iec}{\mbox{i.\,e.\,}}

\newcommand{\egc}{\mbox{e.\,g.\,}}

\newcommand{\mc}[1]{\ensuremath{\mathcal{#1}}}

\newcommand{\ket}[1]{\ensuremath{\left|  #1 \right\rangle}}
\newcommand{\bra}[1]{\ensuremath{\left\langle #1 \right|}}
\newcommand{\bk}[2]{\ensuremath{\left\langle #1 | #2 \right\rangle}}

\newcommand{\matel}[3]{\ensuremath{\bra{#1} #2 \ket{#3}}}
 
\newcommand{\op}[1]{\ensuremath{\widehat{\textsf{\ensuremath{#1}}}}}
\newcommand{\opad}[1]{\ensuremath{\op{#1}^{\dagger}}}

\newcommand{\tr}{\textsf{Tr}}

\newcommand{\be}{\begin{equation}}
\newcommand{\ee}{\end{equation}}
\newcommand{\e}[1]{\mathrm{e}^{#1}}

\begin{document}
\title{Why black hole information loss is paradoxical}\author{David Wallace\thanks{Dornsife College of Letters, Arts and Sciences, University of Southern California; email \texttt{dmwallac@usc.edu}}}

\maketitle

\begin{abstract}
I distinguish between two versions of the black hole information-loss paradox. The first arises from apparent failure of unitarity on the spacetime of a completely evaporating black hole, which appears to be non-globally-hyperbolic; this is the most commonly discussed version of the paradox in the foundational and semipopular literature, and the case for calling it `paradoxical' is less than compelling. But the second arises from a clash between a fully-statistical-mechanical interpretation of black hole evaporation and the quantum-field-theoretic description used in derivations of the Hawking effect. This version of the paradox arises long before a black hole completely evaporates, seems to be the version that has played a central role in quantum gravity, and is genuinely paradoxical. After explicating the paradox, I discuss the implications of more recent work on AdS/CFT duality and on the `Firewall paradox', and conclude that the paradox is if anything now sharper. The article is written at a (relatively) introductory level and does not assume advanced knowledge of quantum gravity.
\end{abstract}

\section{Introduction}

\begin{quote}
Not everyone understands Hawking's paradox the same way[.]

\begin{flushright}
Samir Mathur\footnote{\citeN[p.34]{mathurpedagogical}}
\end{flushright}
\end{quote}

The black hole information loss paradox has been a constant source of discussion in theoretical physics since Stephen Hawking~\citeyear{hawkingbreakdown} first claimed that black hole evaporation is non-unitary and irreversible, but opinions about it differ sharply. The mainstream view in theoretical physics --- especially that part of high-energy physics that has pursued string theory and string-theoretic approaches to quantum gravity --- is that (a) the paradox is deeply puzzling and (b) it must ultimately be resolved so as to eliminate information loss. But prominent critics in physics (\egc Unruh and Wald~\citeyear{unruhwaldpuremixed,unruhwaldinformation}, \citeN{penroseroadtoreality}) and in philosophy (\egc \citeN{belotetalinformationloss}, \citeN{maudlininformation}) seem frankly baffled that anyone could expect information \emph{not} to be lost in black hole evaporation, and often regard the apparent `paradox' as simply the result of confusion. (Maudlin (\emph{ibid}, p.2) goes so far as to suggest that `no completely satisfactory non-sociological explanation' can be given for the paradox's persistence!)

This sharp disagreement arises, I will argue, from an equivocation as to what is meant by `the' information loss paradox. The form most widely discussed in the popular and semi-popular literature, closest in form to Hawking's original discussion, and most directly engaged with by the critics, is based on a clash between apparently-general features of quantum mechanics and the global structure of the spacetime describing a completely-evaporating black hole. \emph{That} version of the paradox is indeed less than compelling: information loss seems prima facie plausible, and in any case the question seems to require a full understanding of quantum gravity to answer and so may be premature.

But there is a second version of the paradox, dating back to \citeN{pagecurve}, which instead rests on the conflict between a statistical-mechanical description of black holes and the exactly-thermal nature of Hawking radiation as predicted in quantum field theory (QFT). \emph{This} version of the paradox is far more compelling, inasmuch as very powerful arguments support both sides in the conflict and yet they appear to give rise to contradictory results. The (mathematical) evidence against information loss advanced by physicists is much more naturally understood in terms of the second version of the paradox, and that version has if anything been sharpened by work in recent years that strengthens both the case for, and the case against, information loss. It remains in the truest sense paradoxical: a compelling argument for a conclusion, a comparably-compelling argument for that conclusion's negation.

The structure of the paper is as follows. In section \ref{background}, I briefly summarise important background facts in the last forty years of black hole physics: the extent to which classical black holes can be given a thermodynamical description; the discovery and significance of Hawking radiation; the progress made in establishing a statistical-mechanical underpinning for black hole thermodynamics.  In section \ref{paradoxes} I present the two forms of the information-loss paradox, focussing on the second and more powerful form. In sections \ref{adscft}--\ref{firewall} I review recent developments (respectively, AdS/CFT duality and the firewall paradox) that bear on this form of the information-loss paradox; section \ref{conclusion} is the conclusion.

Four notes before proceeding. Firstly, I have written this article at about the level of mathematical rigor found in mainstream theoretical physics. I do not attempt full mathematical rigor, which seems premature in any case given the incomplete state of development of the theories being discussed. 

Secondly, for what it's worth my strong impression is that the version of the information-loss paradox I focus on is also the version that high-energy physicists mostly have in mind in their technical work on quantum gravity. However, I do not intend this paper as a historical account and I leave to others the interesting task of disentangling the literature on the topic. 

Thirdly, quantum gravity is a large and varied field. The dominant research program in that field is string theory, and my discussions of quantum gravity in this paper (insofar as they go beyond low-energy regimes which are somewhat better understood) are entirely restricted to the string-theory program. This is partly because the second version of the paradox I discuss has been overwhelmingly developed and discussed in string theory, but mostly on grounds of space and of my own expertise. (I am not an expert on string theory by any means, but I understand the relevant results well enough to have reasonable confidence on how they work and what they rely on; I lack anything like that level of competence in other approaches to quantum gravity.) The reader should not infer anything about the progress, or lack of progress, in understanding black hole statistical mechanics and black hole information loss in other programs from my failure to discuss it here. Readers more sceptical than I about the prospects of string theory, in particular, are warmly encouraged to carry out analogous investigations in other programs. (For loop quantum gravity in particular --- the largest quantum-gravity program outside the string-theory framework --- \citeN{perezloopreview} is a thorough recent review of black hole physics in that program.)

Lastly, unless otherwise noted I work in units where $k_B=c=G=\hbar=1$.

\section{Background}\label{background}

This is a rather brief overview of material I discuss, and critically assess, in much more detail in Wallace~\citeyear{wallaceblackholethermodynamics,wallaceblackholestatmech}. I give only the key references; readers are referred to these papers for more extensive information and references.

\subsection{Black hole thermodynamics}\label{blackholethermodynamics}

In the 1970s it became clear that even classical black holes (that is: black holes as described by classical general relativity, along with phenomenological descriptions of matter fields) behaved in a great many respects as if they were ordinary thermal systems. I review this material in depth in \citeN{wallaceblackholethermodynamics}, and readers are referred there for details and further references, but in summary:
\begin{itemize}
\item Black holes have equilibrium (that is: stationary) states characterised (in their rest frame) only by their energy and by a small number of conserved quantities (charge and angular momentum); `no-hair' theorems \cite{carter1979} establish the uniqueness of these equilibrium states, and perturbations away from equilibrium are damped down quickly \cite[chs.VI-VII]{membraneparadigm}.
\item Small-scale interactions with a black hole (such as lowering charged or rotating matter into the hole) can be divided into `reversible' and `irreversible' in a fashion closely analogous to infinitesimal adiabatic interactions with thermodynamical systems \cite{christodolouruffini}. In this analogy, black hole surface area plays the role of thermodynamic entropy: an infinitesimal change is reversible iff it leaves the area invariant, and no physically-possible intervention can decrease the area. 
\item Stationary black holes of charge $Q$, mass $M$ and angular momentum $J$ satisfy the differential expression 
\be
\mathrm{d}M = \frac{\kappa}{8\pi} \mathrm{d}A - \Omega \mathrm{d}J - \Phi \mathrm{d}Q 
\ee
where $\kappa$ is the surface gravity, $A$ the surface area, $\Omega$ the surface angular velocity, and $\Phi$ the surface electric potential~\cite{bardeenlaws}. This expression can be derived either as an abstract mathematical statement about equilibrium black holes, or as a statement about the small changes in $Q$, $M$, $J$ and $A$ induced by allowing matter to fall into the black hole. This is exactly the expression (under either interpretation) that would encode the First Law of Thermodynamics for a self-gravitating body at thermal equilibrium with that charge, mass and angular momentum, if it had temperature $\lambda \kappa/8\pi$ and entropy $A \lambda$ (for arbitrary positive $\lambda$). Some while later, \citeN{waldnoether} showed how to extend the First Law, and how to define the appropriate generalisation of (entropy $\propto$ area), to arbitrary diffeomorphism-invariant theories of gravity.
\item Hawking's area theorem \cite{hawkingareatheorem} established that no intervention on a black hole could decrease its area, and so extended the (entropy $\propto$ area) idea from infinitesimal to finite processes.
\end{itemize}
The \emph{membrane paradigm} \cite{membraneparadigm} which codified further advances in classical black hole physics in the 1970s and early 1980s, extended this thermodynamic interpretion of black holes to a local description, where a black hole can be regarded (from the perspective of any observer who remains outside) as a thin, viscous, charged, conducting membrane lying at the `stretched horizon', just outside the true event horizon. Under the membrane paradigm, for instance:
\begin{itemize}
\item The increase of black hole area when a charge is dropped onto it can be understood as Ohmic dissipation as the charge flows through the surface;
\item The return to equilibrium of the black hole under the same process can be understood as the spread of the charge from an initially localised region to a uniform charge distribution across the surface;
\item An uncharged black hole rotating in an external magnetic field will develop eddy currents which slow its rotation;
\item Dropping a mass into a black hole induces damped perturbation in its shape, in accordance with the Navier-Stokes equation for viscous fluids.
\end{itemize}
These results, collectively, provide an almost perfect interpretation of a black hole as a thermodynamic system \emph{as long as heat exchanges with other systems are disregarded}. When they are considered, the analogy breaks down entirely as far as classical physics is concerned: the only consistent temperature that can be assigned to a black hole through considerations of thermal contact is zero, and no classically-possible physical process can be reinterpreted as heat flow from a black hole to another thermal system.

\subsection{Hawking radiation}

\citeN{hawking1975} discovered that when quantum field theory (QFT) is applied to a black hole spacetime, it gives rise to field states which, as seen by distant observers, correspond to thermal radiation emitted from the black hole. Hawking radiation (which has since been rederived by many different methods; see \citeN[section 4.2]{wallaceblackholethermodynamics} for a review and discussion of the evidence) can be understood as a consequence of the entanglement of the QFT
vacuum state and of any state locally similar to that state. Field modes on either side of the event horizon are entangled, and from the point of view of an observer who remains outside the event horizon, the effective state of a field mode just outside the horizon can be described by tracing over the physically-inaccessible partner mode just inside the horizon. The resultant state is perfectly thermal as measured by an observer just outside the horizon; because of the black hole's gravitational field, most of this thermal radiation falls back in and the radiation seen by a distant observer differs from perfect black-body radiation by a so-called `grey-body factor' correction.

From our point of view, Hawking radiation has two key features:
\begin{enumerate}
\item It exactly completes the description of a black hole as a thermodynamic system. Hawking radiation from a black hole of surface gravity $\kappa$ has temperature $\kappa/2\pi$, exactly in accord with the black hole temperature contained within the First Law, above (with $\lambda=4$) and it provides a method by which black holes can be put in thermal contact with each other and with other hot bodies (either through radiative transfer or by directly mining the atmosphere of thermal radiation surrounding the hole). 
\item Hawking radiation carries energy (as defined via Noether's theorem applied at large distances) away from the black hole and so can be expected to reduce its mass. Exact calculations remain impossible (they would require a fully-understood theory of quantum gravity) but a variety of different calculations, arguments and numerical simulations within semiclassical gravity (see \citeN[section 4.3]{wallaceblackholethermodynamics} for a review) all give the expected `naive' result that the rate of decrease of the black hole's mass is equal to the Hawking radiation flux at infinity. As such, an isolated black hole will radiate away its mass theoretically until it vanishes entirely, in practice at least until its mass approaches the Planck mass and the assumptions of Hawking's calculation (that full quantum-gravity effects may be neglected) become invalid.
\end{enumerate}

\subsection{Black hole statistical mechanics}\label{black-hole-stat-mech}

So: black holes behave exactly like thermodynamic systems. All \emph{other} thermodynamic systems we know behave that way because their thermodynamics is underpinned by a statistical-mechanical description, so it is natural to speculate that black holes also have a statistical-mechanical description that underlies their thermodynamic behaviour. In particular, the thermodynamic entropy of a statistical-mechanical system is identified with the \emph{microcanonical entropy}, \iec the log of the number of (mutually orthogonal) states available to it, so a statistical-mechanical description for black holes implies that a black hole of given mass, charge and angular momentum, and of area $A$, has $\sim \exp(A/4G)$ (mutually orthogonal) microstates available to it. This speculation was made even before the discovery of Hawking radiation, and by now there is a large amount of calculational evidence supporting it. I review that evidence in depth in \citeN{wallaceblackholestatmech} (other reviews include \citeN{harlowreview} and \citeN{hartmanreview})  but in brief, there are three sources: effective field theory, full quantum gravity, and AdS/CFT duality.

The effective-field-theory route treats general relativity as an ordinary (albeit non-renormalisable) quantum field theory, with the Einstein-Hilbert action as the leading term in an infinite series of interactions and with some unspecified high-energy physics cutting off the divergences in the field-theoretic description at around the Planck length. In this formalism (and temporarily reintroducing the gravitational constant $G$), statistical-mechanical entropy can be calculated using path-integral methods; as was shown by \citeN{gibbonshawking} only a few years after the discovery of Hawking radiation, doing so recovers $S=A/4G$ as the leading-order term. Subsequent work has both clarified the conceptual basis of the calculation and extended it to include interaction terms beyond the Einstein-Hilbert action and higher-order quantum corrections to the leading-order result. The former exactly reproduces Wald's generalisation of the entropy formula; the latter have exactly the required form to renormalise the gravitational constant, ensuring that the $G$ in $S=A/4G$ is the empirically-measured, renormalised Newton constant, not the bare constant that appears in the quantum Lagrangian. (More accurately, the divergent part of the quantum corrections renormalises the constant; the finite part generates additional terms in the entropy formula proportional to $\log M$, which are negligible at classical scales.)

The full-quantum-gravity route depends on one's preferred theory of quantum gravity (and, since we have no fully-worked-out theory of quantum gravity, is necessarily tentative). The most precise (and also the most influential) calculations have been done in string theory, and are restricted to so-called `extremal' black holes; since we will anyway have to consider such black holes later, I pause to give a brief explanation.

Consider first a black hole with nonzero charge $Q$, mass $M$, but no angular momentum. The appropriate black-hole solution to the field equations, the Reissner-Nordstrom solution, only describes a black hole if $|Q|\leq M$ (for larger charges, there is a naked singularity). An extremal black hole satisfies $|Q|=M$ (and is best thought of as the limiting case of black holes closer and closer to extremality). The surface gravity, and thus thermodynamic temperature, of an extremal black hole, is zero; thus they do not radiate and can be thought of as ground states. (In general, a non-extremal black hole with a substantial charge will decay to an extremal black hole rather than evaporating entirely.) This generalises to charged, rotating black holes, as well as black holes in higher dimensions and with more than one sort of charge; in each case, we can describe the black hole by its conserved charge(s) and angular momentum together with the statement that it is extremal.

\citeN{stromingervafa} showed that for a certain class of extremal black hole in five dimensions, the statistical-mechanical entropy can be calculated in string theory; the result, to leading order, exactly matches the area formula. Subsequent calculations have both widened the class of extremal black holes whose entropy can be found in this manner, and refined the calculations to include higher-order corrections and to allow for small perturbations from extremality. The match to the entropy deduced by low-energy methods is exact.

(It's tempting to conclude that (a) this is evidence that string theory is the correct theory of quantum gravity, and/or (b) that the Strominger/Vafa result is only significant \emph{if} string theory is the correct quantum theory of gravity. Both conclusions are too quick. The fact that black hole entropy can be calculated in low-energy quantum gravity --- and, indeed, in QFT on a fixed background --- strongly suggests that any consistent ultraviolet completion of low-energy quantum gravity will reproduce the entropy formula. The match between Strominger and Vafa's result, Gibbons and Hawking's, and the semiclassical prediction, is then evidence firstly that string theory is a \emph{consistent} quantum theory of gravity, and secondly that the sum-over-histories approach to low-energy quantum gravity really is describing the low-energy regime of a consistent theory. See \citeN{wallaceblackholestatmech} for further discussion.)

The AdS/CFT route relies on the conjectured duality (\citeNP{maldacenaconjecture}, \citeNP{gubseretaladscft}, \citeNP{wittenadscft}) between a quantum gravity theory with asymptotically $\AdS_{n+1}\times K$ boundary conditions and a conformal quantum field theory on the $n$-dimensional boundary of $\AdS_{n+1}$, where $\AdS_{n}$ denotes anti-de-Sitter spacetime in $n$ spacetime dimensions and $K$ is some compact space. (Anti-de-Sitter spacetime is most perspicuously thought of here as a sort of box, a covariant version of the `periodic-boundary-condition' boxes often used to make quantum field theories calculationally more tractable.) The conjecture arose in string theory (indeed, arose in part through the extremal-black-hole calculations discussed above) but can be understood, and motivated, as a claim about general quantum-gravity theories on $\AdS$; it cannot be proved formally at present but can be strongly motivated both by physical arguments and by a large number of examples of calculations of the same quantity on either side of the duality, where completely different methods nonetheless give the exact same answer. 

The statistical mechanics of conformal field theories are conceptually under much better control than in the quantum-gravity case, and AdS/CFT duality allows us to calculate entropy and other thermodynamic quantities on the CFT side of the duality and then map them back to the AdS case. The results (which are uncontroversially statistical-mechanical in origin) reproduce the predictions of black-hole thermodynamics; in general qualitatively (\citeNP{wittenadsthermal}, Aharony \emph{et al}~\citeyearNP{hagedorndeconfinementadscft}), but quantitatively and exactly, in those cases where quantitative results can be obtained. (In particular, the extremal-black-hole calculations can be reproduced \cite{stromingerconformal} via AdS/CFT methods.)

In aggregate, these results seem to give strong support to the idea that black hole thermodynamics will be underpinned, in full quantum gravity, by a statistical mechanics of the same general form as that underpinning any other thermodynamic system, and that in particular, any classical stationary black hole  has a large number of microscopic degrees of freedom, in accordance with the statistical-mechanical definition of entropy. The exact form of these degrees of freedom is left somewhat obscure but they appear to have to live in a thin skin around the event horizon, both on general physical grounds (degrees of freedom \emph{within} the event horizon don't seem well-placed to determine a system's thermodynamic properties) and because that seems to be where the detailed calculations place them in those cases where they give any answer at all. A natural way to think of this is as a quantization of the membrane paradigm: not only thermodynamically, but statistically-mechanically, a black hole seems to an outside observer as a thin membrane around the black hole, which has a full unitary description as a quantum system interacting with surrounding radiation, which will be at the Planck temperature as measured by an observer suspended just above its surface, and whose thermodynamical properties are explained in terms of its microscopic degrees of freedom. In particular, in this description Hawking radiation is just ordinary thermal radiation from the surface of the membrane. 

\section{Paradoxes of information loss}\label{paradoxes}

`The' paradox of information loss encompasses a wide range of ideas, but two main versions of the paradox can be discerned. I begin with the version most commonly discussed in the foundational literature, which I review only briefly; for a more detailed review see, \egc, \citeN[ch.7]{waldqft} or \citeN{belotetalinformationloss}. It turns on the violation of unitarity in complete black hole evaporation, and its relation to the causal structure of the evaporating black hole spacetime.

\subsection{Non-unitarity of total evaporation}

\begin{figure}
\begin{tikzpicture}
%infalling matter
\path[fill=gray]  (0,0) to [out=70, in=-90] (1,10) -- (0,10) -- (0,0);
%boundary of diagram
\draw (0,0) -- (10,10);
\draw(10,10) -- (5,15);
\draw(5,15) -- (5,10);
\draw(0,0) -- (0,10);
\draw[decoration = {zigzag,segment length=3mm, amplitude=1mm},decorate] (0,10) -- (5,10);
%region fillers
%\draw[gray](-1,4.5) -- (10,4.5);
%\draw[gray](-1,10.5) -- (10,10.5);
\path[fill=lightgray]  (5,10) -- (10,10) -- (5,15) -- (5,10);
\path[fill=lightgray]  (0,5) -- (10,10) -- (0,0) -- (0,5);
%infalling matter
\path[fill=gray]  (0,0) to [out=70, in=-90] (1,10) -- (0,10) -- (0,0);
\draw[dotted] (0,5) -- (1,5.5);

%event horizon
\draw[dashed](0,5) -- (5,10);
%cauchy surfaces
\draw[black] (0,2.5) to(5,6.25) to (10,10);
\draw[black] (0,7.5) to (5,8.75) to  (10,10);
\draw[black] (5,12.5) to  (7.5,11.25) to (10,10);
%axes
\draw[<->,thick] (0,16) -- (0,0) -- (11,0);
% labels
\node[right] at (5,10) {X};
\node[right] at (4,5.4){$\Sigma_I$};
\node[right] at (4.3,8.3){$\Sigma_{II}$};
\node[right] at (6.2,11.4){$\Sigma_{III}$};
\node[left] at (0,3) {Region I};
\node[left] at (0,8) {Region II};
\node[left] at (5,13) {Region III};
\node[above] at (2.5,10) {Singularity};
\node[right, rotate=90] at (1.2,4) {Infalling matter};
\node[rotate=45] at (2.5,7.5) {Horizon};
\node[left] at (0,16) {t};
\node[right] at (11,0) {r};
\end{tikzpicture}
\caption{Spacetime of a completely-evaporating black hole (angular coordinates suppressed)}
\end{figure}
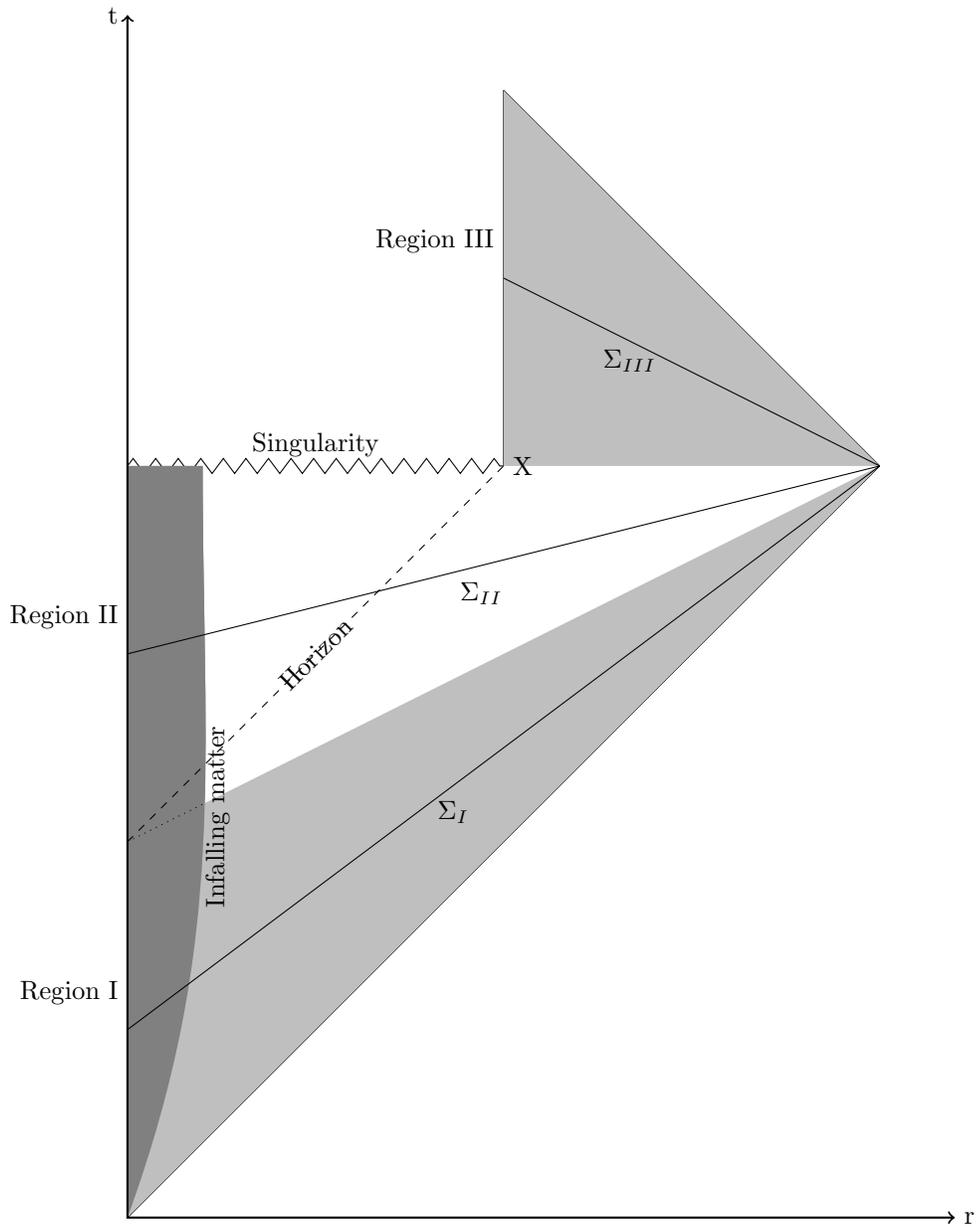

Figure 1 depicts the complete process of black hole evaporation according to the semiclassical description (that is assuming Hawking radiation via QFT on a fixed black-hole background spacetime, and decrease of the mass of that black hole via semiclassical back-reaction). We can distinguish three regions:
\begin{enumerate}
\item The \emph{pre-formation region} $I$ (the lower gray triangle), which is foliated by Cauchy surfaces like $\Sigma_I$.
\item The \emph{evaporation region} $II$ (the white region between the gray triangles), which can be written as the union of $II(int)$ (the region inside the horizon) and $II(ext)$ (the region outside). It is foliated by slices like $\Sigma_{II}$.
\item The \emph{post-evaporation region} $III$ (the upper gray triangle), foliated by slices like $\Sigma_{III}$.
\end{enumerate}
Each region, individually, is globally hyperbolic, as is the combined region $I+II$. The overall spacetime is \emph{not} globally hyperbolic: the point $X$ is a naked singularity.

If QFT on this spacetime accurately describes the black-hole evaporation process, we would expect to be able to describe the physics in region $I \cup II$ by a unitary dynamics, transforming, e.g., a pure quantum state on $\Sigma_I$ to one on $\Sigma_{II}$. The latter state, in general, would be entangled, with the reduced states on $\Sigma_{II} \cap II(int)$ and $\Sigma_{II} \cap II(ext)$ being mixed, but the overall evolution would remain unitary and retrodictable. Since the boundary between $II(int)$ and $II(ext)$ is a future horizon, we would also expect a future-deterministic evolution on $I \cup II(ext)$, so that the reduced state on $\Sigma_{II}\cap II(ext)$ is uniquely determined by the state on $\Sigma_I$, but this evolution will be neither unitary nor past-deterministic (many different initial states would lead to the same black-hole-exterior reduced state.) Strictly, the naked singularity means that there is no well-defined evolution at all from region II to region III, but if we stipulate some well-behaved local physics at the singularity, the evolution from $II$ to $III$ is again deterministic but non-unitary. Indeed, the quantum state on $\Sigma_{out}$ that describes the spacetime immediately upon evaporation is obtained by unitarily evolving the state on $\Sigma_I$ forward to just before the singularity and then tracing out over $II(int)$.

The end result is that the process of black hole formation and evaporation, as described by an observer outside the black hole, is a pure-to-mixed, irreversible transition. The same result can be seen more physically by noting that the quantum state of the exterior in the evaporation region just consists of Hawking radiation, which is (a) perfectly thermal, and hence mixed; (b) determined only by the bulk properties of the black hole (mass, charge, angular momentum) and not by any details of its formation. On slices in region $II$, the full information remains because the formation details are encoded on the state in $II(int)$, but that information is lost once the black hole completely evaporates. (More carefully: that information is not present in any of the slices in region $III$.)

There is a major lacuna in this argument: the final stages of black hole evaporation occur when the black hole's curvature is Planck-scale, and so are well into the regime where semiclassical calculations are unreliable. (This shows up formally in figure 1 via the naked singularity.) But it is hard to imagine filling in the Planck-scale physics in order to save unitarity: the remaining energy does not seem sufficient to encode all the remaining information in a final burst of photons (to say nothing of the question of how it got from the black-hole singularity to the evaporation point $X$); the persistence of the black hole as a Planck-scale `remnant' with fantastically many internal degrees of freedom looks difficult to reconcile with other features of particle physics; the replacement of the point-like naked singularity with a light-like `thunderbolt' seems unmotivated and overkill besides; the removal of the future singularity so that the interior of the black hole forms a `baby Universe' does nothing to save unitarity from the perspective of the external observer.

This powerful argument for non-unitary evaporation is one form of the information-loss paradox, which we can usefully call the `evaporation-time paradox', yet readers might wonder why it deserves to be called a paradox and not just an argument to the interesting conclusion that information is lost. Indeed, that is how \citeN{hawkingbreakdown} originally described it; more recently information loss has been advocated forcefully by Unruh and Wald~\citeyear{unruhwaldpuremixed,unruhwaldinformation}, \citeN[pp.840-841]{penroseroadtoreality} and \citeN{maudlininformation} and more nuancedly by \citeN{belotetalinformationloss}. The stripped-down version of their arguments would be: we have a right to expect unitarity, information preservation, and retrodiction only on globally-hyperbolic spacetimes; the evaporation spacetime manifestly is not globally hyperbolic; so non-unitary evolution is only to be expected.

Arguments for unitarity have been given for this version of the paradox, but frankly they are (to this author) less than compelling. The most straightforward (mostly seen in informal discussion and semipopular work) is simply: unitarity is part of quantum mechanics, so non-unitary evaporation is incompatible with quantum mechanics. But this seems to equivocate on the meaning of `quantum mechanics'; of course we could just \emph{define} quantum physics as incorporating unitary dynamics, but at least some forms of QFT seem perfectly well-defined on non-globally-hyperbolic spacetimes and to have dynamics which is locally unitary but globally non-unitary on those spacetimes. (See the appendix of Belot \emph{et al}, ibid, for discussion; see also \citeN{deutschtimetravel} for an example of the same phenomena in a different context.) Compare: \emph{Hamiltonian} versions of classical electromagnetism are defined only for globally-hyperbolic spacetimes, but classical electromagnetism can also be defined via its local field equations on a much more general class of spacetime. 

More interesting objections come from quantum field theory. In QFT, the amplitude for a transition can generally be expressed as  a sum over all ways in which the transition might come about, which for a full quantum theory of gravity ought to include processes involving formation and evaporation of Planck-scale black holes. Furthermore, high-energy processes like this are typically \emph{not} suppressed in sum-over-histories calculations; rather, their effects can normally be absorbed into the renormalisation of the empirically-measured constants. Prima facie, it looks plausible that non-unitary quantum-gravity effects will make a large difference to low-energy physics and that this difference cannot be normalised away, and some calculations \cite{bankspeskinsusskind,srednickipurity} seem to support this result. But the matter is controversial (see, \egc, \citeN{hawkingvirtualblackholes}, \citeN{unruhwaldpuremixed}, or \citeN{unruhdecoherencedissipation} for dissenting views) and at the least does not seem to provide decisive reasons to reject the argument for information loss, pending a full quantum theory of gravity in which the calculations can be done more carefully.

But while the evaporation-time paradox is the form of the information-loss paradox generally found in popular and foundational literature, it is not the only form. A much more compelling paradox arises when Hawking radiation is considered not just in the light of quantum mechanics in general, but in particular in the light of black hole statistical mechanics.

\subsection{Non-thermality of unitary cooling} 

Following \citeN{pagecurve}, suppose we have some ordinary thermodynamic system with Hilbert space $\mc{H}$, which we want to describe using the microcanonical ensemble. To do so we find some energy interval $\Delta E$, narrow compared to typical energies of the system but wide enough so that the number of energy eigenstates between $E$ and $E+\Delta E$ is large. Then we can define $\mc{H}(E)$ as the subspace spanned by eigenstates of energy with eigenvalues between $E$ and $E+\Delta E$, and write the total Hilbert space as
\be
\mc{H}=\bigoplus_E \mc{H}(E),
\ee
where the sum ranges over energies $E=0,\Delta E, 2 \Delta E, \ldots N \Delta E,\ldots$. The system begins at microcanonical equilibrium at energy $E_0$, and for expository simplicity I assume its initial state is pure (and so is contained in $\mc{H}(E_0)$). It then cools through the emission of thermal radiation: that is, emission of quanta of radiation in highly mixed states. I also assume that the system is large enough that its energy remains in a narrow band as it cools, so that to high accuracy the system's state at any given time is contained within (that is: is a density operator restricted to) some $\mc{H}(E)$.

The original state of the system is pure, and its dynamics are unitary, so the total state of system-plus-radiation is pure. But if the radiation is thermal, each emitted photon will be in a mixed state, and so must be entangled with some other system for the total state to be pure. In thermal radiation no two emitted quanta can be entangled, so each must be entangled with the system. More quantitatively, if the total von Neumann entropy of the  radiation quanta emitted as the system cools from $E_0$ to some lower energy $E$ is $S$, then by unitarity the von Neumann entropy $S_{VN}(E_0,E)$ of the system must also be $S$. 
\begin{figure}
\begin{tikzpicture}
% axes
\draw[<->,thick] (0,6.5) -- (0,0) -- (10,0);
\node[left] at (0,6.5) {Entropy};
\node[right] at (10,0) {t};
% vN entropy
\draw[gray,thick, rounded corners] (0,0) --(4.45,3) -- (8.9,0); 
\node[above, rotate=33] at (2.25,1.5) {von Neumann};
% microcanonical entropy
\draw[dashed](0,6) -- (9,0);
\node[right,rotate=-33] at (2,5) {Microcanonical};
%labels
\draw[thick](4.5,0) -- (4.5,-0.3);
\node[below] at (4.5,-0.3) {Page time};
\draw[thick](9,0) -- (9,-0.3);
\node[below] at (9,-0.3) {Evaporation time};
\draw[thin](4.5,0) -- (4.5,3);
\end{tikzpicture}
\caption{The Page curve}
\end{figure}
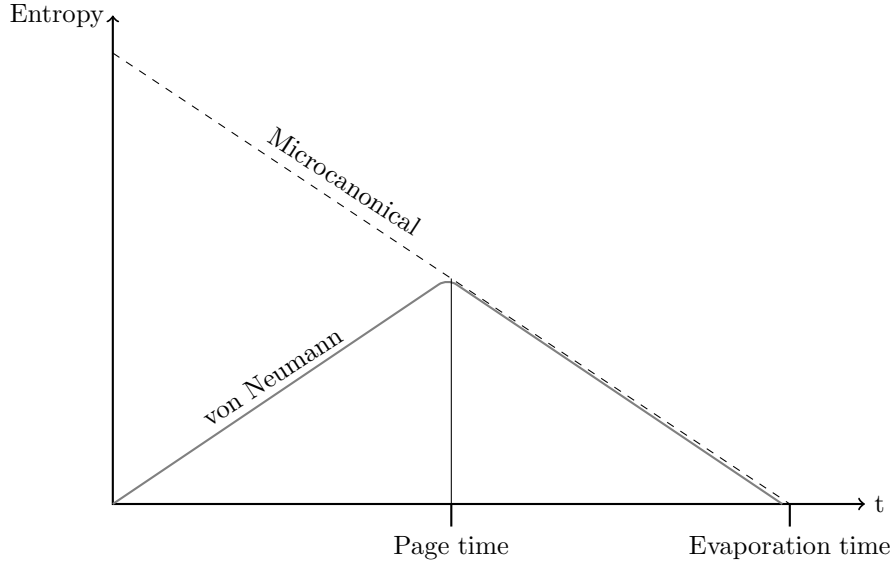
But if the system has energy $E\in [E(t),E(t)+\Delta E]$ at time $t$, its (mixed) state at $t$ must be contained within $\mc{H}(E(t))$, and so must have a von Neumann entropy less than $\log \dim \mc{H}(E(t))$ --- which is to say that the \emph{von Neumann} entropy is bounded above by the \emph{microcanonical} entropy $S_{MC}(E(t))$,
\be
S_{VN}(E_0,E(t))\leq S_{MC}(E(t)).
\ee And the latter (assuming the system has positive temperature) is decreasing as the system cools. There will come a time --- the so-called \emph{Page time}, typically about half way through the cooling process --- when this inequality saturates, and after that point the radiation can no longer be exactly thermal. Instead, the late-time thermal radiation will have to be entangled with the early-time radiation. (Page provides plausibility arguments to the effect that this turnover will be fairly sharp: the radiating body will emit almost-exactly-thermal radiation up to the Page time, so that the von Neumann entropy of the radiating body rises initially until it equals the decreasing microcanonical entropy and then the two remain equal for the rest of the decay process.) The overall process is illustrated in Figure 2: the characteristic time dependence of the von Neumann entropy, initially increasing and then dropping off, is known as the \emph{Page curve}. The initial purity of the system state plays no essential role here provided that the initial state's von Neumann entropy is much lower than its microcanonical entropy. 

According to black hole thermodynamics, black holes are `ordinary quantum-mechanical systems', and the von Neumann entropy of the matter which formed the black hole will be extremely small compared to the black hole's initial thermodynamic entropy. (And, while the stellar precursor of an astrophysical black hole is not plausibly in a pure state, the thermodynamic entropy of such a precursor is typically negligible compared to the entropy of the black hole that forms from it.) The Page time for a Schwarzschild black hole is approximately half the total evaporation time, at which point it will have radiated away about half its mass; after this time, it is not possible for the black hole's radiation to be thermal. Instead, it should be maximally entangled with the early-time radiation (albeit for any physically plausible measurement process this entanglement will be completely undetectable; cf \citeN{harlowhayden}).

The problem is that Hawking radiation --- according to the quantum-field-theoretic calculations --- is \emph{exactly} thermal, displaying no entanglement whatever between early-time and late-time quanta. Indeed, if the calculation is read literally, the successively-emitted quanta are sequentially redshifted down from the trans-Planckian regime at the horizon; each mode of the Hawking radiation is maximally entangled (given its overall expected energy) with a radiation mode inside the event horizon, which has no means of escaping, and so cannot also be entangled with other radiation modes (a principle sometimes called `monogamy of entanglement').

Call this the \emph{Page-time paradox}. (This is basically the form of the paradox as presented in, \egc, \citeN{mathurpedagogical} and \citeN{polchinskiblackholereview}.) It is a clash between the predictions of QFT and the predictions of black hole statistical mechanics that occurs long before complete evaporation, when the black hole is still macroscopic in scale (\iec, there seems no prospect of exotic quantum-gravitational effects coming to the rescue). And while no doubt the exact results of Hawking's calculation need to be modified by various interaction terms and the like, the general form of the calculation seems robust against these modifications and there seems little prospect that these modifications will give the large violations of thermality required to conform to statistical mechanics, at least as long as the basic QFT framework remains intact (see \citeN{mathurpedagogical} for a careful argument to this effect; see also the discussion in section 4.2 of \citeN{wallaceblackholethermodynamics}).

Remnants, or thunderbolts,  or baby universes, no matter how helpful they may be in preserving unitarity, do nothing to preserve the statistical interpretation of black hole entropy or any account of black hole thermodynamics as arising from statistical mechanics in the ordinary way, and so have no role in resolving \emph{this} version of the information-loss paradox. Indeed, I will now show that this version of the information loss paradox can be stated even for situations in which no evaporation occurs at all.

\subsection{Information loss without evaporation}\label{infoparadox-noevap}

For a straightforward realisation of the paradox for non-evaporating black holes, consider the cooling of charged black holes. Recall from our previous discussion that charged non-rotating black holes satisfy the inequality $|Q|\leq M$. Positively-charged black holes preferentially radiate positively-charged particles (and vice versa), but Hawking radiation is dominated by massless particles, so typically the ratio $|Q|/M$ increases in the decay of a charged black hole and it is possible for this inequality to saturate (\iec reach $|Q|=M$). At this point the black hole is extremal; extremal black holes have zero temperature and do not emit Hawking radiation (a good thing for the self-consistency of black hole thermodynamics, since otherwise they would violate $|Q|\leq M$), so the black hole will not decay further.\footnote{This assumes a black hole in a region of spacetime far from any other matter. Astrophysically realistic charged black holes would preferentially absorb oppositely-charged particles (from, \egc, the interstellar medium) at a rate much faster than their evaporation by Hawking radiation, and so would tend away from extremality; thus extremal charged black holes, though a theoretically-possible thought-experiment, should not be understood as representing anything astrophysically realistic.}

Now consider an extremal black hole that ought, according to black hole statistical mechanics, to be in a perfectly mixed state. (Such a black hole could be formed, for instance, by taking a suitably non-extremal black hole and allowing it to approach extremality by decay, such that it reaches the Page time before becoming extremal).  If some (uncharged) radiation is absorbed by the black hole, it will heat up and reradiate the absorbed energy as Hawking radiation. According to QFT, this is a pure-to-mixed transition for physics outside the stretched horizon, since the emitted radiation is perfectly thermal. But according to black hole statistical mechanics, since the quantum state of the black hole is unchanged by the process and the overall interaction is unitary, the outside-the-horizon description should likewise be unitary and the emitted radiation should be in a pure state. (This was an important test case in discussions of the information loss paradox in the 1990s (cf \citeN{maldacenastrominger} and references therein) and will be further discussed in section \ref{near-extremal-unitarity} where we will see that it provides fairly direct evidence for unitary evaporation.)

For a somewhat more complicated case (due to \citeN{maldacena-eternal}), consider a mass-$M$ black hole in a small box, in thermal equilibrium with its atmosphere and with the walls of the box. (One way to implement this is to suppose that the black hole exists in an asympotically $\AdS$ spacetime with effective radius smaller than the black hole's Schwarzschild radius; another \cite{yorkcanonical} is just to impose reflecting boundary conditions on the black hole at a radius less than $1.5 \times$ the Schwarzschild radius). According to the QFT calculations, each emitted photon is in a perfectly thermal state and so is uncorrelated with any other emitted photon. The sharp statement of this in QFT (which is readily demonstrated; see \citeN[section 6.9]{harlowreview}, \citeN{maldacena-eternal} and references therein) is that correlations between field operators at large time separations should fall off exponentially: that is,
\be
C(t)\equiv \langle \op{O}(t)\op{O}(0)\rangle_\rho \equiv \tr (\rho \op{O}(t)\op{O}(0)) \sim \e{-c \beta t}
\ee
where $\rho$ is the thermal state of the black hole atmosphere, $\op{O}(t)$ is some spatially-smeared field operator localised at time $t$, $\beta=8\pi M$ is the inverse temperature, and $c$ is a dimensionless constant. Provided the operator $\op{O}(t)$ is chosen to project onto outward-going states, this should continue to be true, given a QFT description of Hawking radiation, even given the reflecting boundary conditions.\footnote{I am grateful to Gordon Belot for useful discussion on this point.}

But if black hole statistical mechanics is true, then (by the Bekenstein bound; cf discussion in \citeN[section 4.5]{wallaceblackholethermodynamics}) the system of black-hole-plus-box is an ordinary quantum system with a discrete energy spectrum $\op{H}\ket{i}=E_i\ket{i}$. This puts us in the realm to which the Poincar\'{e} recurrence theorem applies and so cannot be compatible with exponentially-decaying correlation functions. To be more precise (here I loosely follow \citeN[section 6.9]{harlowreview}) 
\be 
C(t)=\sum_{i,j}  \frac{\e{-\beta E_i}}{Z(\beta)}|\matel{i}{O}{j}|^2 \e{-it(E_i-E_j)}.
\ee
If we decompose $\op{O}$ into diagonal and off-diagonal parts, $\op{O}=\op{O}_D + \op{R}$, this is
\be \label{correlation}
C(t) = C_D + C_R(t)= \langle \op{O}_D^2 \rangle_\rho + \sum_{i\neq j}\frac{\e{-\beta E_i}}{Z(\beta)}|\matel{i}{R}{j}|^2 \e{-it(E_i-E_j)}.
\ee
(To even give the correlation a chance to drop off exponentially, we should therefore choose an observable with vanishing diagonal part.)
The sums here are over $\sim e^S$ states (where $S$ is the entropy of the ensemble at temperature $\beta$) and $Z(\beta)\sim \e{S-\beta E_0}$ where $E_0$ is the expected ensemble energy. So for this sum to be convergent at small times we can expect $|\matel{i}{R}{j}|^2$ typically to be $\sim \e{-2S}$. For large times, the phase factors $\e{-it(E_i-E_j)}$ can effectively be treated as random, so that $C_R(t)$, very roughly, is a sum of $\e{2S}$ terms of magnitude $\sim \e{-2S}$ and random phase. The theory of random walks predicts that this sum ought to be
\be
C_R(t) \sim \e{-2S} \times \sqrt{\e{2S}} = \e{-S}.
\ee
So if black hole statistical mechanics is true, the correlation factor can initially drop off exponentially, but only until it reaches a value of $\sim\e{-S}$. It will then remain at about that level for a very long time, will occasionally fluctuate back to large values, and eventually, by the quantum version of the Poincar\'{e} recurrence theorem (see \citeN{wallacerecurrence} for a review), will return to its original large value. All this is of course in flat contradiction with the QFT prediction of permanent exponential decay. (The sharp version of the rather heuristic assumptions about $\op{O}$ that I have used in this argument is the \emph{eigenstate thermalization hypothesis} \cite{srednicki-thermalisation}.)

\subsection{The strength of the Page time paradox}

According to one definition~\cite[ch.1]{quineparadox}, a paradox is an apparently-impeccable argument to an impossible conclusion --- such as a pair of apparently-impeccable arguments whose conclusions contradict each other. By this definition, the Page time version of the information-loss paradox is a true paradox. The arguments for black hole statistical mechanics are compelling: quite apart from the general reason to expect a statistical-mechanical underpinning for any thermodynamic system, we have the precise reproduction of the entropy formula (including subleading corrections and renormalisation effects) in low-energy quantum gravity, the equally-precise reproduction of that formula in a large class of extremal black holes using string-theoretic methods, and the recovery of large parts of black hole thermodynamics, qualitatively and quantitatively, from the statistical mechanics of conformal field theory and the AdS/CFT duality. It is unserious to suppose that all of this is simply coincidence. And yet, the arguments from QFT are equally compelling.

Prima facie, there are only two ways forward:
\begin{enumerate}
\item Accept that QFT fails as a description of the entire spacetime of an evaporating black hole; retain the statistical-mechanical underpinnings of black hole thermodynamics; try to understand why and when the QFT description breaks down, given that the breakdown occurs in regimes which prima facie seem `nice' and well within the applicability domain of the theory (\citeNP{polchinskiniceslice}, \citeNP{mathurpedagogical}).
\item Retain QFT, reject black hole statistical mechanics, and find some non-statistical-mechanical understanding of black hole thermodynamics that nonetheless makes the compatibility of black hole and ordinary thermodynamics, and the quantitative results of various statistical-mechanical calculations of black hole entropy, non-miraculous.
\end{enumerate}
Since neither is especially attractive, it's tempting to look for a middle way: to get some understanding of black-hole thermodynamics that holds onto its statistical-mechanical underpinnings but permits non-unitary decay, and/or to find a way to preserve at least most of the QFT description of the evaporation process even while allowing for long-time entanglement between Hawking quanta (violating (4), for the black hole at equilibrium). But results over the last twenty years have sharpened the paradox and virtually foreclosed on the possibility of a middle way, as AdS/CFT duality has provided direct mathematical evidence for duality and as the firewall paradox has made stronger and more explicit the violation of QFT in the statistical-mechanical description.

\section{Evidence for unitarity from AdS/CFT}\label{adscft}

AdS/CFT duality, briefly mentioned in section \ref{background}, has become a growth industry in theoretical physics since its initial discovery and even a cursory discussion of its structure and the evidence for it would double the length of this paper. (I give a brief overview from the point of view of black hole physics in \citeN{wallaceblackholestatmech}; more detailed reviews include Aharony \emph{et al}~\citeyear{aharonyadscftreview}, \citeN{harlowreview}, \citeN{hartmanreview}, and \citeN{kaplanadscft}; for philosophical discussion see, \egc, \citeN{deharogaugegravity} or \citeN{tehholography}.) Here I take the existence of the duality mostly for granted and simply discuss its implications for the unitarity of black hole decay. Specifically, in sections \ref{near-extremal-unitarity}--\ref{eternal-unitarity} I reprise the two `non-evaporating' forms of the paradox discussed in section~\ref{infoparadox-noevap}; in section \ref{unitarity-of-decay} I give a direct argument from AdS/CFT and from Poincar\'{e} recurrence that full evaporation is a unitary process.

\subsection{Unitarity of Hawking radiation from near-extremal black holes}\label{near-extremal-unitarity}

As I noted in section \ref{black-hole-stat-mech}, the statistical-mechanical entropy of certain extremal black holes, calculated via string theory, exactly matches the predictions of black hole thermodynamics. If such a black hole absorbs a small amount of uncharged mass, it will be perturbed away from equilibrium and acquire a temperature, and the leading-order change in entropy in this process again matches the thermodynamic prediction \cite{horowitzstrominger}. 

More strikingly, and more relevantly for our purposes, \citeN{maldacenastrominger} were able to reproduce the exact \emph{spectrum} of the perturbed black hole's Hawking radiation. To expand: recall (section~\ref{blackholethermodynamics}) that a radiating black hole is not  a black body when observed far from the horizon: radiation emitted in the near-horizon regime has some probability to be scattered back across the horizon, and that probability depends on the radiation's angular momentum and its energy. These \emph{grey-body factors} cannot be calculated analytically for the Schwarzschild black hole, but they can be for the near-extremal black hole analysed by Maldacena and Strominger, giving the emission rate for photons of energy $\omega$ as a --- reasonably complicated --- function of $\omega$ and the parameters characterising the black hole. Maldacena and Strominger found this function, and then found the equivalent function for the decay rate of the excited string state corresponding (in \citeN{stromingervafa}'s analysis) to the black hole, through string perturbation theory. They match exactly, despite the completely different calculational tools applied in the two cases (solving the wave equation on the black hole spacetime in one case, calculating emission rates for a dilute gas of string excitations in the other).

This provides pretty powerful support for the hypothesis that black hole decay is a unitary process. To quote Maldacena and Strominger (emphasis in original),
\begin{quote}
The string decay rates, extrapolated to the large black hole region, agree precisely with the semiclassical Hawking decay rates in a wide variety of circumstances. However, the string method not only supplies the decay \emph{rates}, but it also gives a set of unitary \emph{amplitudes} underlying the rates. We find it tempting to conclude that these extrapolated amplitudes are also correct. It is hard to imagine a mechanism which corrects the amplitudes, but somehow conspires to leave the rates unchanged.

The robust nature of the string picture is very significant because it allows us to directly confront the black hole information puzzle \ldots According to Hawking information is lost as a large excited black hole decays to extremality. On the other hand the string analysis \ldots gives a manifestly unitary answer.
\end{quote}

In retrospect, the Maldacena-Strominger calculation can be recognised as a form of AdS/CFT correspondence, realising the duality between the near-horizon region just outside a near-extremal black hole (which can be approximated as $\AdS_3 \times K$ for compact $K$) and two-dimensional conformal field theory on the boundary of that spacetime. In either case the system is coupled to degrees of freedom describing the far-from-horizon region; the CFT description of that coupling is manifestly unitary and exactly reproduces the AdS results for overall scattering levels. See \citeN[pp.114--130]{hartmanreview} for a presentation that makes the AdS/CFT aspect explicit. (And note that as such, the calculation relies only on AdS/CFT duality and not on any specific features of string theory.)

\subsection{Unitarity of Hawking radiation for large black holes}\label{eternal-unitarity}

Radiation from near-extremal black holes is one of the two no-evaporation forms of the information loss paradox I discussed in section \ref{infoparadox-noevap}. The other --- Maldacena's eternal black hole --- can also be analysed via AdS/CFT methods. Recall that for a black hole in a box, large enough to be in stable equilibrium with its atmosphere, QFT calculations of Hawking radiation contradict black hole unitarity for sufficiently large times: the former predict that long-time correlation functions decay exponentially without limit, while unitarity (in a finite box) predicts that the correlations reach a minimum value of $\sim e^{-S}$, and that they eventually undergo Poincar\'{e} recurrence back to their original values. Since a large AdS black hole is a `black hole in a box, in equilibrium with a radiation bath', we can use AdS/CFT duality to work out the correlation functions (the operators whose correlations are calculated are well outside the horizon, in a region where the translation between AdS and CFT descriptions is fairly well understood). We get these results:
\begin{enumerate}
\item For comparatively short times, the exponential decay of the correlations can be recovered on the CFT side of the correspondence \cite{pr-infalling}.
\item For very long times, the discreteness of the spectrum of the CFT Hamiltonian guarantees that the exponential decay ceases, and that the correlation coefficients display the behaviour predicted by unitarity, in contradiction with the predictions of QFT applied to the interior. \cite{maldacena-eternal}
\end{enumerate}
So the AdS/CFT correspondence, applied to large black holes, provides further `compelling evidence' \cite[p.92]{harlowreview} that black hole decay is unitary.

\subsection{Unitarity of black hole decay via AdS/CFT}\label{unitarity-of-decay}

In the case of black holes in AdS space which evaporate completely, it is tempting to conclude immediately that that decay is unitary: after all, the AdS description is dual to a CFT description that is manifestly unitary. But this has been challenged as too quick (cf \citeN{unruhwaldinformation}, \citeN{maudlininformation}), so here I give a more explicit argument. 

Specifically, suppose we have a QG theory on asymptotically $\AdS$ space (with the usual reflecting boundary conditions normally assumed in AdS/CFT duality), and a CFT on the boundary of that space, with Hilbert spaces $\mc{H}_G$, $\mc{H}_{CFT}$ respectively. We choose a foliation $\Sigma_t$ of $\AdS$ compatible with the time translation symmetry map on $\AdS$ (so that the boundary of that foliation defines a foliation for the CFT, and write $\op{H}$ for the Hamiltonian of the CFT with respect to that foliation. Then we make the following assumptions (working in the Heisenberg picture):
\begin{description}
\item[A. Perturbative QG sector:] For any time $t$ and any state of the interior describable semiclassically as an excitation $\psi$ of the vacuum (by, say, gravitons or matter particles) on $\Sigma_t$ which nowhere is dense enough to form an event horizon, there is a state $\ket{\psi;t}\in\mc{H}_G$ which represents that state in the full quantum-gravity theory. Call the subspace spanned by such states $\mc{H}_{G,P}\subset \mc{H}_G$.
\item[B. CFT spectrum:] The spectrum of \op{H} is discrete and bounded below, and $\op{H}$ is at most finitely degenerate.
\item[C. Perturbative duality:] There is a unitary map $\op{V}$ from $\mc{H}_{G,P}$ into $\mc{H}_{CFT}$, such that 
\be 
\bk{\psi;t}{\psi';t+\tau}=\matel{\psi;t}{\opad{V}\exp(-i \tau \op{H})\op{V}}{\psi;t}
\ee
(that is, the map between the interior and boundary theories commutes with the dynamics in at least the perturbative sector.)
\end{description}
(A) is a fairly minimal requirement of representational adequacy for the quantum-gravity theory; (B) is a standard result about conformal field theories on compact spaces~(see, \egc, the discussion in \citeN{harlowreview}); (C) is a small fragment of full AdS/CFT duality. (The map $\op{V}$ in (C) can be constructed fairly explicitly, at least to first order in perturbation theory; see \citeN{harlowstanfordadscft}.)

Now let $\psi$ be an excitation of the AdS vacuum corresponding to a large amount of diffuse infalling matter at time $t$ which will at a later time (with very high amplitude) form a black hole. Given (A), there is a state $\ket{\psi,t}$ that represents this excitation.
Given (B), the quantum version of the Poincar\'{e} recurrence theorem applies to the boundary CFT (see \citeN{wallacerecurrence} for a review) and so for any perturbative state $\ket{\psi,t}$ we can find a time $T$ such that to an arbitrarily good approximation,
\be
\exp(-i T\op{H})\op{V}\ket{\psi,t}\simeq\op{V}\ket{\psi',t}.
\ee
Given (C), $\ket{\psi,t}\simeq \ket{\psi,t+T}$; in other words, after the recurrence time, the bulk theory will describe a multiparticle state containing all the information of the pre-black-hole state, and no information has been lost. Of course, $T$ is vastly longer than the black hole's decay time, but if the evolution is unitary all the way forward to $T$, in particular it is unitary during the decay process.

\section{The firewall paradox}\label{firewall}

AdS/CFT duality seems to have persuaded most of the high-energy physics community (notably including \citeN{hawkingrecant}) \emph{that} black hole evaporation is unitary. But the duality remains poorly understood as far as the black hole interior is concerned, and the question remains: \emph{how} is it unitary, in the face of the clear arguments from QFT for information loss. In the last few years this question has become more urgent, with deep problems emerging in the hitherto-leading strategy for reconciling the exterior and interior descriptions. Section \ref{complementarity} reviews that strategy, normally called `black hole complementarity'; section \ref{firewall-subsection} reviews the so-called `firewall paradox' that appears to invalidate it.

\subsection{Black hole complementarity and the black hole interior}\label{complementarity}

At first sight, there is a conflict between black hole statistical mechanics and QFT that is much more direct than the information-loss paradox.  After all, QFT predicts that an observer will encounter nothing special as they fall freely across the horizon from a starting point high above the black hole, and in particular will measure radiation that deviates only weakly from empty space. This seems hard to reconcile with the description of that same infalling observer that will be given by a fiducial (\iec, hovering) observer: for that second observer, whose observations are confined to the outside-horizon region, the quantum black hole is represented (cf discussion in section~\ref{black-hole-stat-mech}) by a membrane just above the horizon, whose local temperature is order the Planck temperature, and the infalling observer will collide with that membrane and  rapidly be thermalised by it.

The \emph{semiclassical} membrane paradigm demonstrates that this is far too quick. In that paradigm (recall) a quantity of charge dropped onto a black hole will spread out uniformly over the horizon, with a known timescale, generating heat as it does so through Ohmic dissipation. This description of stretched-horizon physics is formally compatible with --- indeed, is derived from --- an underlying physics in which the charge drops smoothly through the horizon and continues towards the singularity. The point is simply that there is a mathematically valid description of the physics of the black hole exterior in terms of the stretched horizon; the metaphysical question of whether that description is \emph{true} is interesting but somewhat tangential.

\citeN{susskindthorlaciusuglum} proposed extending this idea --- that interior information is encoded in information about surface perturbations --- from semiclassical physics to the full quantum theory of the black hole. Just as in the classical case, the proposal is that the selfsame physical process can be described in terms of coordinates adapted to the interior of the black hole, or in terms of the degrees of freedom of a membrane just outside the event horizon. A sketch of this idea for the infalling observer would go as follows. The fall of the observer from infinity onto the stretched horizon can be described equally well with respect to stationary observers hovering above the horizon at a fixed height, and with respect to inertially falling observers (and that description is analysable in terms of known physics; cf \citeN{unruhwald1982}). Since the inertial description tells us that the observer does not catch fire and burn up during this part of their journey, the stationary description must give the same result, so that the observer's passage through the black hole atmosphere is uneventful despite its increasingly high temperature. (The intuitive feeling that this can't happen physically can be assuaged by noting that the observer passes through the hotter part of the atmosphere at extreme relativistic speed and so interacts with the atmosphere for a very short proper time; note that an observer whose fall begins from quite close to the horizon and so takes a much longer proper time to fall in will encounter highly blue-shifted Hawking radiation right from the start of the fall, even in an inertial-frame description, and so will be consistently described as being burned up in both descriptions.)

When the observer reaches the stretched horizon, according to black hole statistical mechanics they will rapidly become thermalised (specifically, thermalisation takes time $\sim M \log M$, as can be read off the thermodynamics of the membrane paradigm). But `thermalisation' is a coarse-grained notion: it means that \emph{for any observable relevant to the exterior physics}, the expectation value of that observable is the same  when calculated with the `thermalised' state as with the true microcanonical-equilibrium state (the projector onto the energy eigensubspace of the black-hole Hilbert space). This is perfectly compatible with observables relevant to the black hole \emph{interior} having expectation values that deviate sharply from the values calculated from the microcanonical-equilibrium state, and indeed which describe the infalling observer in accordance with general relativity.

\citeN{susskindthorlaciusuglum} called this duality of surface and interior descriptions \emph{black hole complementarity}. The name invokes the non-commutativity of quantum-mechanical observables: just as the same physical process can be described with respect to a basis of definite-position or definite-momentum states and may look very different in the two descriptions, so the horizon-crossing process can be described with respect to a basis appropriate to exterior physics or one appropriate to the infalling observer's situation. (Regrettably, ``complementarity'' \emph{also} invokes Bohr's somewhat obscure philosophy of meaning --- and Susskind \emph{et al} explicitly refer to that philosophy --- but so far as I can see it is not essentially required in black hole statistical mechanics.)

However, the parallel with semiclassical complementarity is imperfect, precisely because of the (Page time) information loss paradox: after the Page time, QFT continues to predict exactly thermal radiation, whereas black hole statistical mechanics requires that radiation to be entangled with early-time radiation, and so if black hole statistical mechanics is correct then QFT must actually be \emph{wrong}, not just a redescription of the same physics, at late times. As we shall now see, the firewall paradox of \citeN{firewall}, based on earlier ideas by \citeN{mathurpedagogical}, strongly suggests that this failure of QFT is not simply some kind of long-range effect but completely invalidates the semiclassical description of the black hole interior.

\subsection{Firewalls}\label{firewall-subsection}

The firewall paradox can be stated in various more-or-less precise ways (see, \egc, \citeN{boussofirewall}, \citeN[section 7]{harlowreview}, \citeN[section 6]{polchinskiblackholereview}, \citeN{susskindfirewall},  as well as the original sources above), but in essence it works like this. Consider some photon mode $B$ (more accurately: a wavepacket concentrated on some such mode) describing photons emitted well after the Page time. That mode will be in a thermal state at the appropriate-time black hole temperature. We now have two apparently-contradictory claims:
\begin{enumerate}
\item According to a QFT description, $B$ is fully entangled with some mode $\tilde B$ just inside the event horizon.
\item According to black hole statistical mechanics, $B$ is fully entangled with the early-time radiation emitted by the black hole.
\end{enumerate}
No quantum state can be fully entangled with two different systems (sometimes called `monogamy of entanglement'), so this seems close to a contradiction. We might hope to finesse this by remembering the idea of complementarity --- that the same underlying physics can be described in radically different ways, so that the system has one valid description where $B$ is entangled with an interior mode, and one where it is entangled with early-time radiation. But as Almheiri \emph{et al} point out, in principle (though not in practice; cf \citeNP{harlowhayden}) an observer could
\begin{enumerate}
\item Collect all the radiation emitted from the black hole up to the Page time;
\item Carry out a complicated operation on that radiation to distil a single photon mode $C$ that is fully entangled with $B$;
\item Linger close to the event horizon, and perform a joint measurement on $B$ and $C$ (more precisely, on many such pairs $B_1,C_1 \ldots B_N,C_N$) to verify their entanglement;
\item Jump into the black hole.
\end{enumerate}
Assuming that the observer's own local physics can be consistently described by quantum mechanics, they can't consistently find $B$ to be entangled with $\tilde B$; indeed, $B+\tilde B$ must be in a product state. This directly contradicts the QFT assumptions underpinning Hawking radiation, and so undermines the basis of at least some of the arguments for Hawking radiation which got black hole statistical mechanics going in the first place. (Anecdotally it seems to be a matter of dispute in the physics community to what extent the derivation of Hawking radiation is undermined by the firewall argument.) More dramatically, complete distentanglement of QFT modes across the event horizon corresponds to a Planck-scale wall of energy at the horizon --- the `firewall' --- that seems physically inexplicable and quite at odds with the general-relativistic idea of an event horizon as a globally-defined, locally-inaccessible phenomenon. 

The firewall paradox is only five years old at time of writing, and it would be premature to try to summarise, far less assess, the wide variety of different responses that have been offered (see \citeN[section 8]{harlowreview} for a partial review). But it has roiled the community, disrupted what had been a fairly solid consensus in favour of black hole complementarity, and thrown the theory of the black hole interior wide open.

\section{Conclusion}\label{conclusion}

\begin{quote}
Thus we find ourselves in the enviable situation of having an interesting problem with no really satisfying answer; if we are lucky this means that we will learn something deep.

\begin{flushright}
Daniel Harlow\footnote{\citeN[p.117]{harlowreview}.}
\end{flushright}
\end{quote}

The black hole information paradox, understood in its most powerful form, is a clash between the unitary description of Hawking radiation implied by statistical-mechanical models of the black hole horizon, and the non-unitary description given by quantum field theory. It is neither a foolish failure by physicists to appreciate the subtleties of non-globally-hyperbolic spacetimes, nor something harmlessly resolved by AdS/CFT duality (the latter, at most, gives us reason to expect the ultimate resolution to be unitary). It is a deep puzzle arising from enormously-plausible yet apparently-contradictory lines of reasoning within quantum gravity, and at present it is completely opaque how it is to be resolved.

This is a good thing. There are very few obvious empirical clues as to the nature of quantum gravity; in their place, the best we have are the highly demanding and often unexpected consistency constraints given by the internal structure of lower-energy physics. We have learned much about the form of any satisfactory quantum theory of gravity by trying to satisfy those constraints; as and when we find a satisfactory resolution to the information-loss paradox, we will have learned still more.

\section*{Acknowledgements}

I am grateful to Gordon Belot, Sean Carroll, Erik Curiel, Eleanor Knox, James Ladyman, and Tim Maudlin for useful discussions, and in particular to Jeremy Butterfield for detailed and careful comments on a draft version. I'm also grateful to three anonymous referees for careful and helpful comments.

%\bibliography{../bib/general2}
%\bibliographystyle{chicago}

\end{document}